\documentclass[fleqn,usenatbib,useAMS]{mnras}

\usepackage{graphicx}	% Including figure files
\usepackage{amsmath}	% Advanced maths commands
\usepackage{amssymb}	% Extra maths symbols
\usepackage{multicol}        % Multi-column entries in tables
\usepackage{bm}		% Bold maths symbols, including upright Greek
\usepackage{pdflscape}	% Landscape pages

\usepackage[T1]{fontenc}
\usepackage{ae,aecompl}

\usepackage{newtxtext,newtxmath}
\title[Timing Galaxy Mergers]{Galaxy Mergers in UNIONS -- II: Predicting Timescales in the Post-Merger Regime}
\author[L. Ferreira et al.]{
Leonardo Ferreira$^{1,}$$^{2}$\thanks{Contact e-mail: \href{lferreira@uvic.ca}{leonardo.ferreira.furg@gmail.com}},
    Sara L. Ellison$^{1}$,
    David R. Patton$^{3}$,
    Shoshannah Byrne-Mamahit$^{1}$,
    \newauthor
    \ Scott Wilkinson$^{1}$,
    Robert W. Bickley$^{1}$\\
$^{1}$Department of Physics \& Astronomy, University of Victoria, Finnerty Road, Victoria, BC V8P 1A1, Canada\\
$^{2}$Instituto de Matemática Estatística e Física, Universidade Federal do Rio Grande, Rio Grande, RS, 96203-900, Brazil\\
$^{3}$Department of Physics and Astronomy, Trent University, 1600 West Bank Drive, Peterborough, ON K9L 0G2, Canada}

\date{Accepted 2026 January 12. Received 2026 January 09; in original form 2025 December 03.}

\pubyear{2023}

\begin{document}
\label{firstpage}
\pagerange{\pageref{firstpage}--\pageref{lastpage}}
\maketitle

\begin{abstract}
    Galaxy mergers are critical events that influence galaxy evolution by driving processes such as enhanced star formation, quenching, and active galactic nucleus (AGN) activity. However, constraining the timescales over which these processes occur in the post-merger phase has remained a significant challenge. This study extends the MUlti-Model Merger Identifier (\textsc{Mummi}) framework to predict post-merger timescales ($T_{PM}$) for galaxies, leveraging machine learning models trained on realism-enhanced mock observations derived from the IllustrisTNG simulations. By classifying post-merger galaxies into four temporal bins spanning 0 to 1.76 Gyr after coalescence, \textsc{Mummi} achieves time classification accuracies exceeding 70 per cent. We apply this framework to the Ultraviolet Near Infrared Optical Northern Survey (UNIONS), yielding a catalog of 8,716 post-merger galaxies with $T_{PM}$ predictions and stellar masses $\log(M_*/M_\odot) \geq 10$ at redshifts 0.03 < z < 0.3. These results provide a robust methodology to connect galaxy interaction timescales with physical processes, enabling detailed studies of galaxy evolution in the post-merger regime.

\end{abstract}


%%%%%%%%%%%%%%%%%%%%%%%%%%%%%%%%%%%%%%%%%%%%%%%%%%

%%%%%%%%%%%%%%%%% BODY OF PAPER %%%%%%%%%%%%%%%%%%

% The MNRAS class isn't designed to include a table of contents, but for this document one is useful.
% I therefore have to do some kludging to make it work without masses of blank space.
\begingroup
\let\clearpage\relax
\endgroup
\newpage
\section{Introduction}

Galaxy mergers are among the most transformative events in the life cycle of galaxies. They profoundly alter stellar, gaseous, and dynamical structures, triggering a cascade of processes including enhanced star formation rates \citep{scudder2012, patton2013, Stierwalt2015, Violino2018, Garay-Solis2023}, Active Galatic Nuclei (AGN) activity \citep{ellison2011, Ellison2019, Satyapal2014, Bickley2023, Bickley2024, Byrne-Mamahit2023, Byrne-MamahitB}, quenching mechanisms \citep{Springel2005, Hopkins2008, ellison2022, Wilkinson2022}, and morphological changes that can persist for billions of years \citep{Toomre1977, Lotz2008, Conselice2014Review, Wilkinson2024, Ferreira2024a}. These interactions also contribute to the hierarchical assembly of galaxies, a cornerstone of the $\Lambda$CDM cosmological model \citep{Bluementhal1984, Duncan2019, Duan2024, Patton2024}.

Despite their central role, the timescales over which these processes occur remain poorly constrained. While pre-merger interactions are well characterized through pair statistics, including their correlation with separation and velocity \citep{Patton2000, Ellison2008, Mantha2018, Duncan2019, Duan2024}, the post-merger regime lacks precise temporal constraints \citep{Ellison2013, Pawlik2016, Bickley2022}. Morphological disturbances, commonly used as indicators of recent mergers together with traditional methods, yield only limited insight into the time elapsed since coalescence \citep{NairandAbraham2010, Privon2013, Wilkinson2024}. As a result, studies of post-mergers often treat them as a single, temporally averaged population, which can mask the rich evolutionary trends within this phase \citep{Ellison2013, Li2023, Bickley2023}. 

To address this lack of observability, machine learning frameworks have emerged as promising tools for classifying galaxies and identifying mergers, and they also show potential for disentangling merger timescales \citep{Walmsley2022, Cheng2023, Walmsley2023}. One particularly successful approach involves mapping large-scale cosmological simulations to the observational domain to train deep learning models, thereby creating a bridge between simulations and observations \citep{Pearson2019, Wang2020, Alexandra2020, Ferreira2020, Bottrell2019, Bickley2021, Alexandra2021, Ferreira2024a, Schechter2025} and encoding simulation information that can be leveraged at inference time \citep{ERGOML}.

Recently, \citet{Koppula2021}, \citet{Pearson2024}, and \citet{Pearson2025} used simulated galaxies from HorizonAGN and IllustrisTNG to demonstrate that time predictions can be performed using only a galaxy’s visual features. These studies frame the problem as a regression task by estimating the time to or since coalescence based on how many simulation snapshots lie between the current galaxy’s state and its coalescence moment. Combined with the time-resolution limitations of modern large-scale cosmological simulations, this uncertainty raises questions about the applicability of these models to real galaxies.

We move towards measuring timescale robustly using the Multi-Model Merger Identifier (\textsc{Mummi}) framework \citep{Ferreira2024a}, extending it to estimate post-merger timescales. \textsc{Mummi} combines Convolutional Neural Networks (CNNs) with Vision Transformers (ViTs) to achieve high-accuracy classification of mergers into pre- and post-coalescence categories, leveraging both simulation and observational datasets. In its initial application to the Ultraviolet Near Infrared Optical Northern Survey (UNIONS)  survey \citep{UNIONS2025}, \textsc{Mummi} identified over 13,000 high-confidence galaxy mergers, enabling detailed studies of merger-driven processes \citep{Ferreira2024a}.

In the work presented here, we discuss how we use \textsc{Mummi} to predict the time since coalescence ($T_{PM}$) for post-merger galaxies, a measure that enables tracking the evolution of physical processes in the post-merger regime.  Applying these timescale predictions to the UNIONS survey enables the creation of the first detailed observational timeline of post-merger evolution, encompassing trends in star formation \citep{Ferreira2024b}, quenching \citep{Ellison2024b}, AGN activity \citep{Ellison2025}, and stellar mass build-up \citep{Ellison2026}.

This paper is organized as follows: \S~\ref{sec:data} describes the data and methodology for training and validating the extended \textsc{Mummi} framework, including the use of mock observations and ensemble voting. \S~\ref{sec:RESULTS} presents the performance of the $T_{PM}$ predictions. Finally, \S~\ref{sec:summary} summarizes our findings and outlines future directions for extending this work to higher redshifts and more complex datasets.
 
We assume the same cosmological model used by IllustrisTNG, which is consistent with the \citet{Planck2016} results that show $\Omega_{\Lambda,0} = 0.6911$, $\Omega_{m,0} =  0.3089$, and $h = 0.6774$.

\section{Data \& Methods} \label{sec:data}

In this work, we describe how we extended the \textsc{Mummi} framework \citep{Ferreira2024a} to estimate post-merger timescales from imaging, enabling the tracking of physical processes over time in the post-merger regime. This new step operates on top of the existing merger classification in \textsc{Mummi}, acting only on galaxies already identified as post-merger candidates.

We use a subset of the same mock imaging dataset constructed from IllustrisTNG galaxies \citep{Rodriguez-Gomes2015, Pillepich2018, Nelson2018}, incorporating UNIONS realism (see Section 2 of \citealt{Ferreira2024a} for full details).  In this section, we briefly present \textsc{Mummi} and the post-merger galaxies used to train the machine learning ensemble for predicting post-merger timescales. 

\subsection{The Optical Imaging from UNIONS}\label{subsec:CFIS}

We apply \textsc{Mummi} to the UNIONS survey $r$-band images. The $r$-band imaging data were obtained with the 3.6\,m Canada–France–Hawaii Telescope (CFHT) on Maunakea, covering $4\,861\ \mathrm{deg}^2$ in the $u$ and $r$ filters and reaching a $5\sigma$ surface-brightness limit of $28.4$\,mag\,arcsec$^{-2}$ \citep{UNIONS2025}.

Each $r$-band pointing consists of three single-exposure visits separated by small dithers.  This observing pattern both refines the astrometric/photometric solutions and fills the inter-CCD gaps of the MegaCam mosaic.  Raw frames are processed with the \textsc{MegaPipe} pipeline \citep{Gwyn2008,Gwyn2019}, where bias subtraction and night-sky flat-fielding are applied.  Astrometric registration employs \textit{Gaia}~DR2 \citep{GaiaCollaboration2018}, while Pan-STARRS 3$\pi$ $r$-band photometry \citep{Chambers2016} provides differential and absolute zero points.  The calibrated exposures are co-added onto a regular grid of $0.5^{\circ}\times0.5^{\circ}$ tiles, using PS1 stars as in-field flux standards.  The final stacks attain a typical 5$\sigma$ point-source depth of $r=25$\,mag, median seeing of $\sim0.7''$, and a pixel scale of 0.187$''$.

Our models are then issued on a cross-matched UNIONS Data Release~5 (DR5) with spectroscopic galaxies from SDSS DR7 \citep{Abazajian2009}.  We restrict the redshift range to $0.01<z<0.3$: the lower bound ensures that galaxies fit within the MegaCam field of view, whereas the upper bound minimises resolution-driven losses and contamination by unresolved quasars.  This procedure yields 235\,354 galaxies with secure redshifts. From these, we have previously identified 42,764 galaxy mergers \citep[around $13,000$ high-confidence][]{Ferreira2024a}.

\subsection{Post-merger Selection in TNG100-1}\label{subsec:selection}

The selection of post-merger galaxies is based on the merger tree data from the IllustrisTNG100-1 simulation \citep[e.g with \textsc{subfind}][]{Rodriguez-Gomes2015}. We define post-mergers as galaxies that have undergone a merger event within the past 11 snapshots in the simulation ($\sim1.75~\mathrm{Gyr}$) with a stellar mass ratio of $\mu \geq 0.1$, and that do not have close companions within a projected distance of $50~\mathrm{kpc}$. This ensures that the selected systems are dynamically distinct and isolated from any ongoing pair interactions. To further mitigate contamination from spurious mass measurements caused by numerical artifacts, we measure stellar masses when the merging galaxies are at least $50~\mathrm{kpc}$ apart, thus avoiding biases introduced by mass exchange or stripping during the coalescence process \citep{Rodriguez-Gomes2015, Patton2020, Byrne-MamahitB}.

\begin{figure*}
    \centering
    \includegraphics[width=0.95\linewidth]{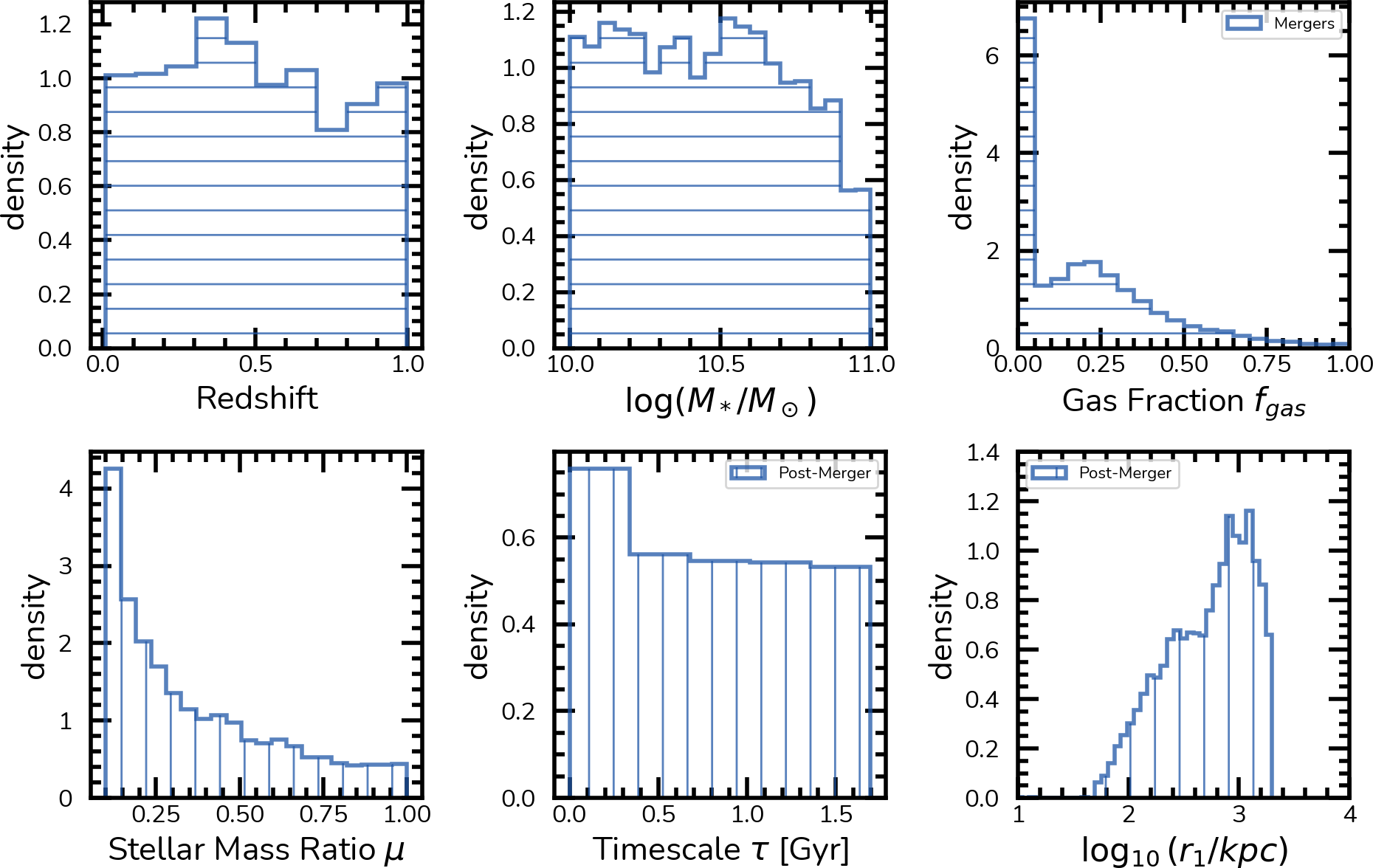}
    \caption{\textbf{IllustrisTNG post-merger sample statistics}. We show histograms for the physical properties of our mock sample of UNIONs-like simulated galaxies. Redshift, stellar masses and gas fractions are shown above, while stellar mass ratios, true time since latest merger event and the separation of the closest companion in 3D space in the simulations can be found on the bottom row. Our post-merger sample displays a wide range of properties, from wet to dry mergers, encompassing minor mergers and major mergers.   }
    \label{fig:sample_stats}
\end{figure*}

The post-merger sample is restricted to galaxies with stellar masses $M_\ast > 10^{10}~\mathrm{M_\odot}$, ensuring a minimum particle resolution that is sufficient to produce realistic morphologies. From the initial TNG100-1 pool, these criteria yielded a total of $21,485$ post-mergers, covering a redshift range of $z = 0$ to $1.0$. Moreover, we adopt the same upper stellar mass limit of $10^{11} M_\odot$ from \cite{Ferreira2024a}. Additionally, to produce a balanced dataset with respect to timescales, we undersample each of the eleven snapshot-specific subsets to match the size of the snapshot immediately after the merger event, which is the sparsest subset. In practice, this means that for every later snapshot we randomly select, without replacement, the same number of unique mergers as available in the post-coalescence snapshot. This yields a total of $17,897$ post-mergers. In Figure \ref{fig:sample_stats}, we display the overall properties of the post-merger sample. The top row shows physical information such as redshift, stellar mass, and gas fractions, while the row below presents merger-specific properties, including stellar mass ratio, post-merger timescale, and closest companion 3D separation (up to 2 Mpc, see \citealp{Patton2020}) in the simulation box.

\begin{figure*}
    \centering
    \includegraphics[width=0.99\linewidth]{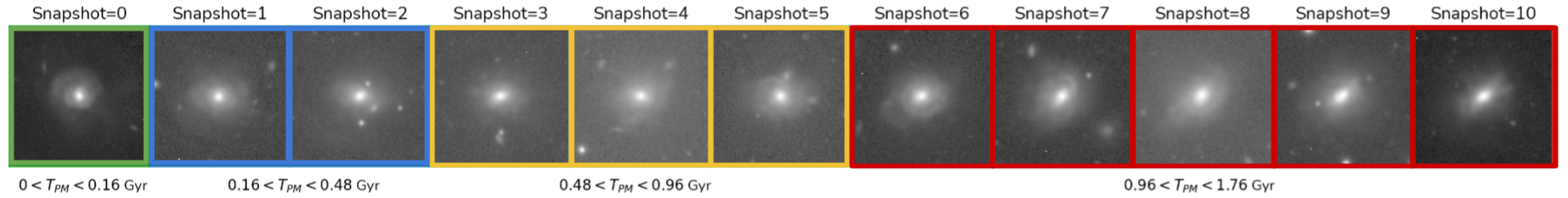}
    \caption{\textbf{Snapshots of simulated post-merger galaxies from IllustrisTNG, highlighting the temporal evolution of morphological features across four post-merger time bins}. Each snapshot corresponds to a representative galaxy classified in the respective bin: 0 < $T_{PM}$ < 0.16 Gyr (green outline), 0.16 < $T_{PM}$ < 0.48 Gyr (blue outline), 0.48 < $T_{PM}$ < 0.96 Gyr (yellow outline), and 0.96 < $T_{PM}$ < 1.76 Gyr (red outline). Early post-mergers exhibit prominent morphological disturbances, such as tidal tails and asymmetries, which diminish in intensity with time. By the final bin, galaxies appear largely relaxed, reflecting the gradual fading of merger-induced features.}
    \label{fig:example}
\end{figure*}

In Figure \ref{fig:example}, we show mock images for a single galaxy in the simulation (same collection of particles) across all 11 snapshots, corresponding to a $\sim1.75~\mathrm{Gyr}$ time evolution. Our final classifications will be ultimately based on four bins rather than 11 (\S \ref{sec:RESULTS}), denoted by different colours and time ranges at the bottom of the figure.

{ 
\subsubsection{Numerical Definition of Coalescence and Resolution Effects}

A critical aspect of timing galaxy mergers in simulations is the numerical definition of "coalescence", which in the IllustrisTNG merger trees is intrinsically linked to the subhalo identification logic of the \textsc{subfind} algorithm \citep{Springel2011}. In \cite{Ferreira2024a} we give a detailed overview of how we select galaxy mergers from the merger trees, and how we define their time of coalescence. However, here we discuss potential numerical issues that can arise from the subhalo finding algorithm itself and how we try to mitigate these effects.

To identify substructures, \textsc{subfind} first locates host haloes using a standard Friends-of-Friends (FoF) algorithm with a linking length of $b=0.2$ times the mean inter-particle separation. Within each FoF group, local overdensities are identified by estimating the density field using an adaptive kernel interpolation based on the nearest neighbours of each particle \citep{Springel2011}. The algorithm identifies local density maxima and progressively lowers a density threshold to grow candidates until they meet a saddle point connecting them to a structure with a higher density peak. Once these potential subhalo candidates are defined by their isodensity contours, an unbinding procedure is applied: particles with positive total energy (kinetic plus potential) relative to the subhalo's centre are iteratively removed. This ensures that the final subhalo consists only of gravitationally bound particles. A critical feature of \textsc{subfind} is its exclusive particle assignment, where each particle is assigned to a single subhalo (or the background "fuzz" of the host), which can lead to the artificial truncation of satellite properties in dense environments where tidal fields are strong.

However, this "moment of merger" is a discrete approximation of a continuous physical process and is subject to several numerical subtleties that can decouple simulation timing from observational intuition. For example, the timing can be sensitive to numerical resolution, as higher-resolution simulations generally allow subhaloes to be tracked for longer periods because they resist tidal stripping, potentially delaying the recorded coalescence time compared to lower-resolution counterparts \citep{Onions2012}. Furthermore, finder convergence remains a challenge; configuration-space finders like \textsc{subfind} may "lose" a subhalo near the dense central regions of a host galaxy earlier than phase-space finders might, particularly when the secondary's particle count drops below a specific threshold, typically $\sim$20--32 particles \citep{Knebe2011}. These effects are further complicated by baryonic physics, as the presence of a concentrated stellar core—such as those found in our $\log(M_{*}/M_{\odot}) \ge 10$ sample -- increases the binding energy of the subhalo, making it more resilient to tidal disruption and allowing \textsc{subfind} to track the system deeper into the post-merger regime than in dark-matter -- only simulations \citep{Dolag2009}.

The definition of a "merger" is further complicated by the fact that different subhalo identification algorithms can yield significantly different results, even for well-resolved systems. \citet{Forouhar2025} found that while different algorithms generally agree on the properties of isolated field haloes, their predictions increasingly diverge in dense environments such as the centres of host haloes. The "moment of coalescence" -- when a finder ceases to track a subhalo -- is dominated by the algorithmic definition of boundedness rather than purely physical disruption. 

To mitigate these effects and ensure our results align with morphological evolution rather than numerical artifacts, we define our merger properties, such as the stellar mass ratio $\mu \ge 0.1$, when the progenitors are at least 50~kpc apart \citep{Byrne-MamahitB}. This procedure ensures that the physical characteristics of the interaction are established before the final stages of numerical coalescence, where the algorithm's ability to distinguish substructure becomes most uncertain. Additionally, we make sure that all our galaxies correspond to subhalos with a cosmological origin \citep{Pillepich2018}, avoiding subhalos generated from baryonic effects, such as disk fragmentation. Furthermore, given that IllustrisTNG snapshots are sparse ($\sim 162$ Myr), we assume that timing issues are of this order or lower, and mostly impact pre-merging galaxies. Thus, the impact is minimal in our dataset, and limited to our closest to coalescence snapshot, where close pairs might be mislabeled as post-mergers, showing up as two separated nuclei (\ref{subsec:contaminants}). 

}
\subsection{MUlti-Model Merger Identifier -- \textsc{Mummi}}

\textsc{Mummi} is a supervised deep learning framework designed to classify galaxy mergers and distinguish their evolutionary stages. This hybrid architecture combines CNNs \citep{AlexNet} and ViTs \citep{visionTransformer}, leveraging the strengths of both approaches to achieve strong classification accuracy and purity, even for complex morphological features across Gyr-long timescales \citep{Ferreira2024a}.

\textsc{Mummi} is trained using a dataset of synthetic galaxy images derived from the IllustrisTNG100-1 cosmological simulation. The dataset includes 6.4 million mock images generated using the \texttt{RealSimCFIS} pipeline \citep{Bottrell2017b, Bickley2021}, which reproduces the observational properties of the UNIONS r-band \citep{UNIONS2025} images. These mock observations incorporate realistic effects, including redshift dimming, point-spread-function (PSF) convolution, and background contamination from real UNIONS sky images. Galaxies in the sample span $10^{10} M_\odot < M_* < 10^{11} M_\odot$ in stellar mass, $0.0 < z < 1.0$, and a range of merger stages, including interacting pairs up to 11 snapshots ($\sim1.75~\mathrm{Gyr}$) before coalescence, post-mergers up to 11 snapshots ($\sim1.75~\mathrm{Gyr}$) after coalescence, and control galaxies matched in physical properties but lacking recent or imminent mergers.

The hybrid architecture of \textsc{Mummi} integrates CNNs, specifically EfficientNet models, for efficient feature extraction, and SwinTransformers, which capture the long-range spatial dependencies critical for identifying tidal features. The framework employs an ensemble of models trained on distinct subsets of the dataset to ensure diversity and uses a consensus-based voting system, inspired by Condorcet’s theorem, to maximize classification purity \citep{Boland1989}. On top of that, \textsc{Mummi} is also a hierarchical framework, with sequential steps that feed into each other, narrowing down the specificity of the question at hand. In \cite{Ferreira2024a} we discuss in detail Step~1 and Step~2, but here we provide a brief summary (the reader is referred to the original paper for full details). The first step of \textsc{Mummi} is comprised of an ensemble of 20 models (10 CNN-ViT pairs), trained specifically to separate mergers from non-mergers. \textsc{Mummi}'s first step achieves a purity of 95\% even for mergers with Gyr-long timescales. Sequentially, the second step has a single pair of models trained to classify the merger stage, assigning the galaxy pre-coalescence (pair) or post-coalescence (post-mergers) status. Step~2 models are trained only with galaxy mergers, with no controls. At this task, \textsc{Mummi} achievesa success rate of 97\%. This hierarchical approach effectively reduces the false-positive rate by 75\% compared to previous machine learning classifiers \citep{Ferreira2024a}.

In the work presented here, we present the third \textsc{Mummi} step. It follows sequentially only those post-mergers detected in \textsc{Mummi}'s Step~2, with the goal to determine how much time has passed since the last merging event. This is done with a new ensemble of 20 models (again, 10 pairs of models) -- very similar to the first step -- that classify post-mergers according to the number of snapshots that have passed since the merger event. Specifically, these models are trained to predict up to a maximum of 11 simulation snapshots, resulting in timescale predictions in the range $0 < T_{PM} < 1.76$ Gyr, as these are separated by roughly $\sim 160$ Myr each in TNG100-1. It is important to note that the correspondence between snapshots and lookback time is not perfectly uniform: the cumulative lookback time covered by 11 snapshots varies for the redshift interval $0<z<1$, with a dispersion of $18$--$20$~Myr around the nominal 160~Myr between consecutive snapshots \citep{Nelson2018}. However, this uncertainty is negligible for our final predictions given our confidence intervals are always wider in lookbacktime than this time difference. Additionally, every galaxy is re-imaged at 20 redshifts in the training set, forcing the models to marginalize it during training.

In practice, Step~3 of \textsc{Mummi} uses a new ensemble of 20 networks composed in equal numbers of \texttt{EfficientNetB0} CNNs and \texttt{SwinT} Transformers\footnote{We use the \texttt{Tensorflow} implementation available at \url{ https://github.com/shkarupa-alex/tfswin}}. Each model takes as input $256\times256$ pixel images and outputs a probability vector for the 11 snapshot classes. We adopt the \texttt{Lion} optimizer \citep{LionOpt} with a cosine annealing learning rate schedule, oscillating between $10^{-4}$ and $10^{-5}$ over 200 epoch cycles, including a warming up period of 10 epochs and early stopping. Most models achieve convergence in a single cycle.

Each of the 17,897 post-merger galaxies selected in \S~\ref{subsec:selection} is augmented in 4 different orientations and 20 artificial redshifts, resulting in 1,431,760 total images \citep[see Appendix A in][for full details]{Ferreira2024a}. These data are then split in $80\%$ for training, while $20\%$ is used for validation and tests. We take care to do this split on a galaxy by galaxy basis so that no viewing angle and redshift of a given system can appear in more than one subset (all realizations are either in the training set or validation set, never both). The training set is sectioned into 10 subsamples, each training a pair of CNN-SwinT. Details on sample selection and total training size are discussed in \S~\ref{subsec:selection}. Further data augmentation on the fly such as rotations, horizontal/vertical flips, random translations, zooms, and random occlusion are included, preventing the models from memorizing particular tidal morphologies, keeping the CNNs and ViTs exposed to comparable stochasticity. We refer the reader to Appendix A in \cite{Ferreira2024a} for more details.

%Unlike Steps 1 and 2 of \textsc{Mummi}, Step 3 aggregates the predictions from each model as probability distributions across the 11 snapshots and then reduces them based on their mode, generating confidence intervals that inform the final predictions and assess their robustness. 

\section{RESULTS}\label{sec:RESULTS}

\begin{figure}
    \centering
    \includegraphics[width=0.95\linewidth]{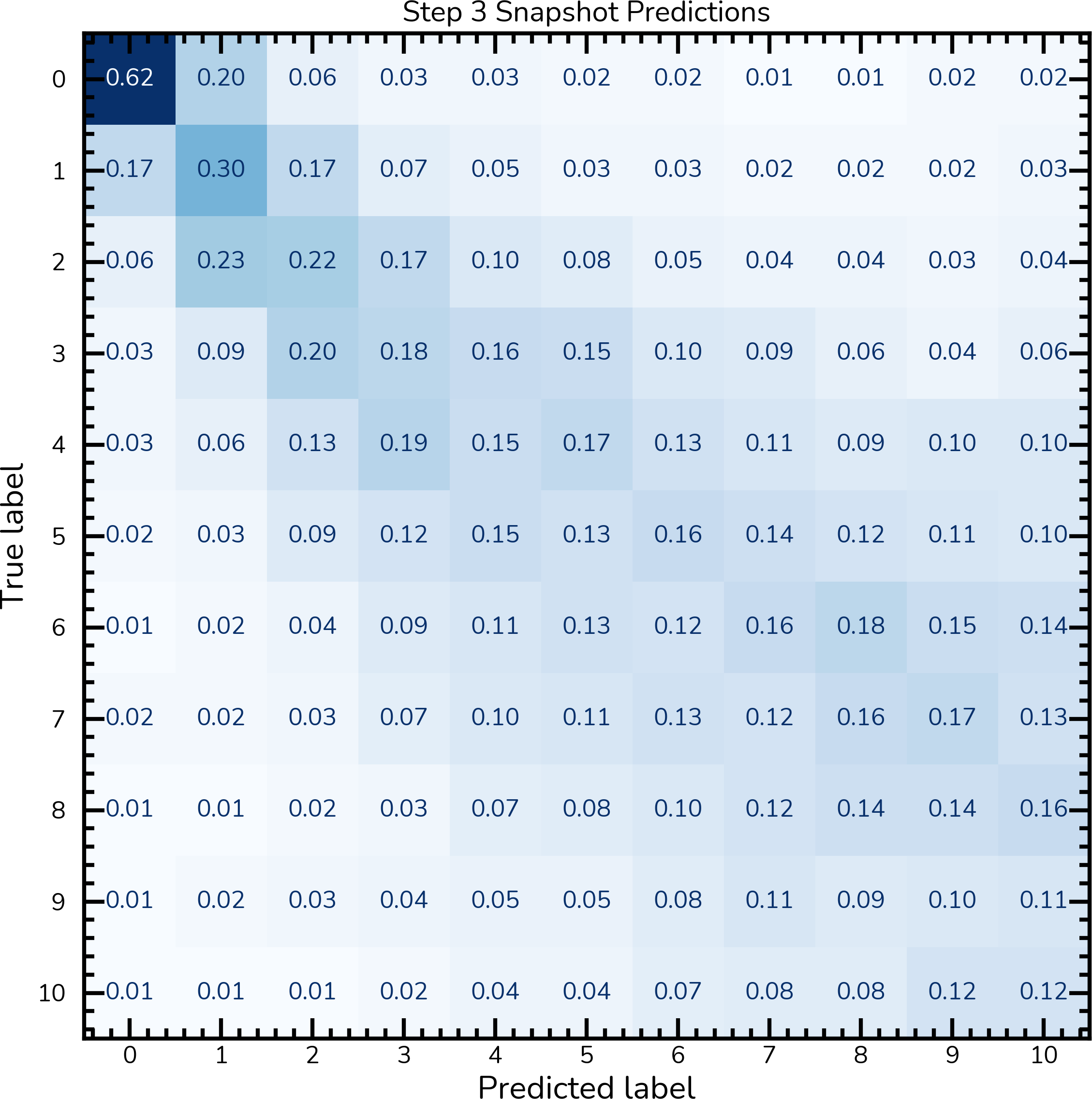}
    \caption{\textbf{Confusion matrix showing the performance of \textsc{Mummi} Step 3 in predicting the number of simulation snapshots that have passed since the merger event}. The true labels (vertical axis) and predicted labels (horizontal axis) correspond to the number of snapshots elapsed. Diagonal elements represent the fractions of galaxies correctly classified for a given snapshot count, while off-diagonal elements indicate misclassifications. The matrix highlights \textsc{Mummi}'s strong performance for early snapshots (e.g., snapshot 0 with 84\% accuracy) and intermediate snapshots, while later stages (e.g., snapshots 8–10) show increasing confusion, reflecting the gradual fading of merger-induced features and the difficulty in distinguishing relaxed systems.}
    \label{fig:CMFULL}
\end{figure}

In this section, we discuss the performance of \textsc{Mummi} timescale predictions. First, we evaluate the ensemble of models in Step 3, trained to predict how many simulation snapshots have passed since the merger event (\S \ref{subsec:snap}). Additionally, we describe how we convert these predictions into four distinct timescale bins to track the temporal evolution of post-mergers in both simulations and observations (\S \ref{subsec:timescalepredictions}). In \S~\ref{subsec:qualityflags}, we show how we leverage the ensemble statistics to create a statistical robustness criterion that enhances model performance. Subsequently we explore how \textsc{Mummi} deals with misclassifications in \S~\ref{subsec:contaminants}. Moreover, we explore potential biases and selection effects of applying the timescale quality control in \S~\ref{subsec:bias}. Finally, we discuss the results of applying \textsc{Mummi} timescale predictions to the UNIONS survey (\S~\ref{subsec:UNIONSresults}).

\subsection{Snapshot Predictions}\label{subsec:snap}

We first evaluate \textsc{Mummi}'s ability to predict how many snapshots have passed since coalescence on a reserved test set. The test set was already processed with Step 1 and Step 2 of \textsc{Mummi}, ensuring that Step 3 is evaluated under conditions analogous to real observations. In particular, we exclude any post-merger galaxy not selected by \textsc{Mummi} in the earlier steps, even if it is flagged as a post-merger in the simulation.

Secondly, we recast the task as a \emph{classification} rather than a pure regression problem, assigning every simulation snapshot to its own class.  This choice is motivated by several practical and astrophysical considerations:

\begin{enumerate}
    \item \textbf{Discrete ground‐truth.}  The simulation delivers coalescence times at a finite set of snapshot epochs; treating the labels as categorical avoids interpolating between non-existent intermediate outputs and respects the intrinsic time resolution of the data. This mitigates the need for higher time resolution simulations \citep[e.g][]{Pearson2024, Pearson2025}.
    \item \textbf{Robustness to label noise.}  Small uncertainties in the exact moment of coalescence (e.g.\ due to the finite snapshot cadence) translate into class boundary migrations instead of large numerical residuals, reducing the impact of noisy labels on the loss function.
    \item \textbf{Degeneracy management.}  Different merger histories can converge to similar morphologies at late stages; a binned approach tolerates such degeneracies by clustering look-alike galaxies into the same class instead of forcing a single-valued continuous target.
\end{enumerate}

Consequently, each model outputs a probability distribution over all classes. We combine the ensemble predictions by averaging them and selecting the snapshot corresponding to the peak of the averaged distributions\footnote{One can use the mode peak from all model outputs as the snapshot prediction; however, it is less meaningful when combined with the confidence intervals later. Results using the mode are similar.}. The statistics of the averaged probability distribution will be used later to define confidence intervals that exclude uncertain or noisy predictions, thereby enhancing the ensemble’s overall performance (see \S~\ref{subsec:qualityflags}).

In Figure \ref{fig:CMFULL}, we show the confusion matrix based on the mode predictions for the 11 snapshots evaluated in the test set. The confusion matrix quantifies the accuracy of classifications across 11 snapshot bins in the aggregated ensemble predictions, where each bin corresponds to the number of snapshots elapsed post-merger. The diagonal elements represent the fraction of galaxies correctly classified into their respective bins, whereas the off-diagonal elements indicate misclassifications.

The matrix in Figure \ref{fig:CMFULL} shows that \textsc{Mummi} achieves strong performance for early snapshots, especially at snapshot 0, where the classification accuracy reaches 62\%. This high accuracy reflects the distinct morphological disturbances that characterize galaxies immediately after coalescence. Intermediate stages, such as snapshots 3 to 5, exhibit moderate classification accuracy, as they feature overlapping morphological properties that can lead to confusion with adjacent bins. This trend is consistent with the gradual fading of merger-induced features, making it more challenging to distinguish older post-merger systems. Later snapshots (6+) show increased confusion spread.

The off-diagonal elements further reveal a pattern of progressive misclassification into neighbouring bins. For example, galaxies at snapshot 1 are occasionally misclassified as snapshot 0 or 2, illustrating the gradual morphological transition over time. Despite this, extreme misclassifications, such as assigning late-stage galaxies (e.g., snapshot 10) to early bins, are rare, underscoring the robustness of the \textsc{Mummi} framework. Even though a random classifier with 11 classes would have an average accuracy of 9\% in each bin, the accuracy in each snapshot bin remains too low for a robust statistical assessment of the predicted times at this time resolution.

\subsection{Timescale Predictions}\label{subsec:timescalepredictions}

Given that misclassifications are preferentially placed into adjacent time bins, we leverage this behaviour to improve the predictive power of our models. Instead of predicting the time post-merger in the original 11, equally spaced, snapshots we define timescale bins of variable width that maximize our ability to distinguish different timescales across the post-merger regime, encompassing the snapshot bins with higher misclassification rates. We restructure the problem to classify galaxies into one of four time bins based on the time elapsed since coalescence ($T_{\rm PM}$), given that the time difference between each snapshot is approximately $\sim 160$ Myr on average for the $0 < z < 1$ range. Thus, we use the following definition\footnote{Various other configurations were tested, but we settled on this one to maximize final performance.}:

\begin{itemize}
    \item Bin 1: $0  \ \rm Gyr < T_{\rm PM} < 0.16$ Gyr (1 snapshot, immediate);
    \item Bin 2: $0.16 \ \rm Gyr < T_{\rm PM} < 0.48$ Gyr (2 snapshots, short);
    \item Bin 3: $0.48  \ \rm Gyr< T_{\rm PM} < 0.96$ Gyr (3 snapshots, intermediate);
    \item Bin 4: $0.96  \ \rm Gyr< T_{\rm PM} < 1.76$ Gyr (5 snapshots, long);
\end{itemize}

These bins are chosen to capture broad key phases of morphological evolution following coalescence: from the immediate coalescence ($0 \ Gyr < T_{\rm PM} < 0.16 \ Gyr$), when the stellar masses have just merged centrally, to the early post-coalescence stage ($0.16 \ Gyr < T_{\rm PM} < 0.48 \ Gyr$), when coalescence is fully complete but morphological disturbances remain strong, followed by an intermediate period ($0.48 \ Gyr < T_{\rm PM} < 0.96 \ Gyr$), during which physical processes begin to settle, 
and finally a late stage ($0.96 \ Gyr < T_{\rm PM} < 1.76 \ Gyr$), when the impact of the merging event has largely relaxed and physical processes have returned closer to a non-merger state. These four phases allow us to trace the evolution of star formation enhancements \citep{Ferreira2024b}, quenching \citep{Ellison2025}, AGN \citep{Ellison2024b}, and stellar mass build-up \citep{Ellison2026} in observations during the post-merger regime. 

We use the mean snapshot time difference ($\sim 160 \ \rm Myr$) as a way to convert these bins into lookback time. However, given that this snapshot separation varies with redshift in the simulations, taking the mean for this conversion can result in discrepancies accumulating up to a snapshot difference (in time) depending on the galaxy redshifts. These differences, however, are greater for longer timescales (more snapshots), and thus are negligible given the width of our longer timescale bins. On top of that, every galaxy in our training samples is re-imaged in 20 redshifts, thus this is also marginalized during training (e.g a z=1 galaxy is re-observed at 20 distinct redshifts) \citep{Ferreira2024a}. Moreover, we remind the reader that predictions made directly into single snapshot domains in lookback time are not useful given the poor performance presented in Figure \ref{fig:CMFULL}.

\begin{figure}
    \centering
    \includegraphics[width=0.95\linewidth]{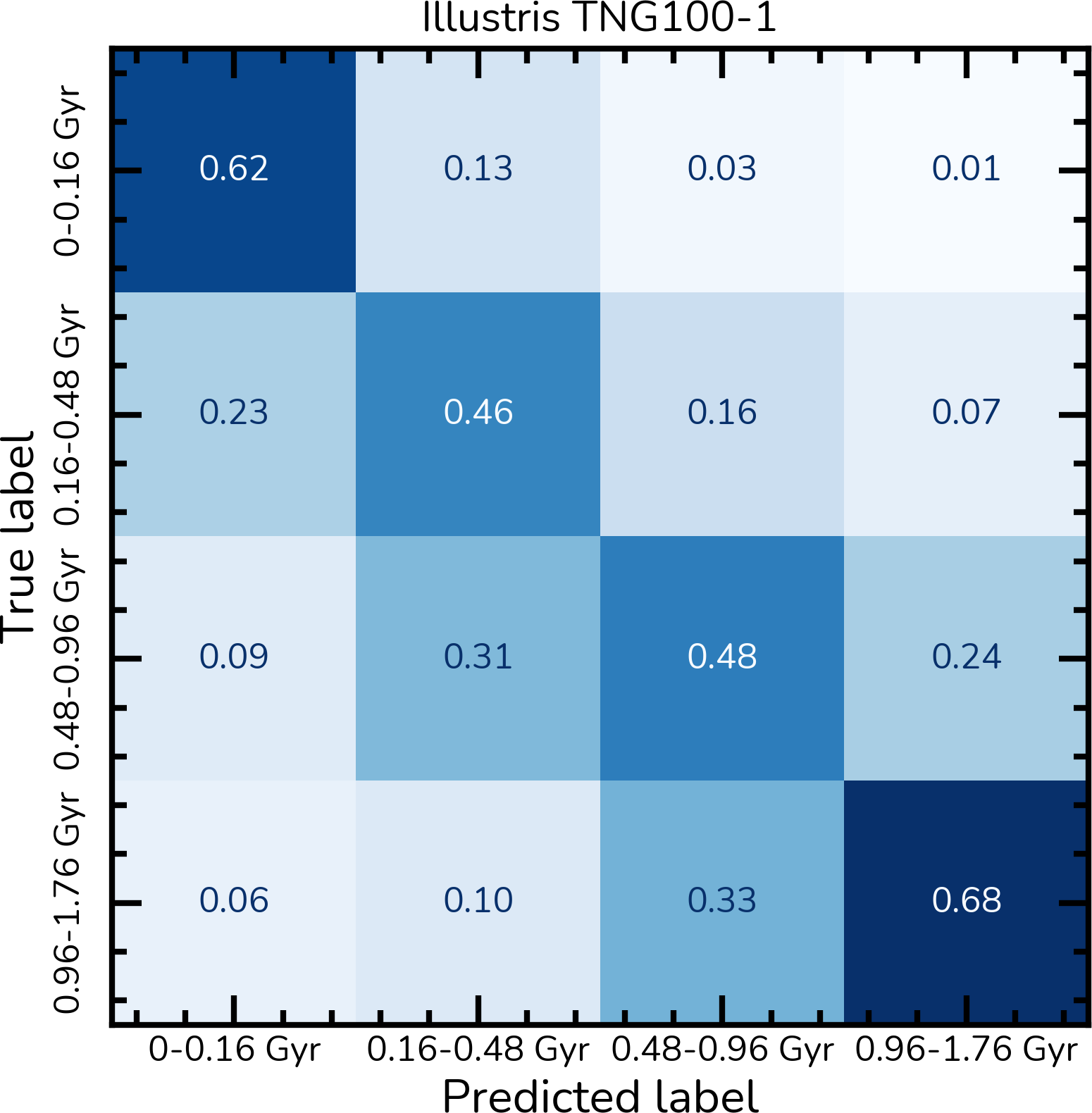}
    \caption{\textbf{Confusion matrix showing the performance of \textsc{Mummi} before the application of probability flags in predicting post-merger timescales across four temporal bins}: $0 \ \rm Gyr < T_{PM} < 0.16 \ \rm Gyr$, $0.16 \ \rm Gyr < T_{PM} < 0.48 \ \rm Gyr$, $0.48 \ \rm Gyr< T_{PM} < 0.96 \ \rm Gyr$, and $0.96 \ \rm Gyr < T_{PM} < 1.76 \ \rm Gyr$. Each cell represents the fraction of galaxies with a true temporal bin label (vertical axis) assigned to a predicted bin (horizontal axis). Diagonal elements highlight correctly classified galaxies, with classification purity ranging from 47\% to 68\%. Misclassifications (off-diagonal elements) are more prominent in intermediate bins, reflecting the challenges of distinguishing transitional stages. This matrix provides a baseline for evaluating the improvements introduced by subsequent probability flagging. For ease of comparison with Figs.~\ref{fig:CMFULL} and \ref{fig:cm-small}, all confusion matrices in this work are normalized to show purity in the diagonal elements.}
    \label{fig:cm-small-before}
\end{figure}

With this new binning structure, we re-evaluate the performance of \textsc{Mummi} in predicting post-merger timescales using the same test set described in the previous section. Figure \ref{fig:cm-small-before} summarizes the classification accuracy across the four temporal bins. Each cell in the matrix represents the fraction of galaxies with a given true label that are assigned to a particular predicted label. The diagonal entries represent correct classifications, whereas the off-diagonal entries indicate misclassifications. Overall, the new confusion matrix demonstrates improved performance in disentangling different time slices across the post-merger regime, with classification accuracies ranging from 68\% to 46\%. The highest accuracy is achieved for the long timescale bin. The immediate post-mergers ($0 < T_{PM} < 0.16$ Gyr) is identical to the first bin in the larger matrix due to the fact that it represents a single snapshot regardless of the other bins.

Notably, the misclassification rates remain asymmetric across the temporal bins. However, the overall performance remains limited, reflecting the rarity of post-mergers in the real universe. In the next section, we explore how to use the statistics from our ensemble of models to exclude uncertain and spurious classifications that are more likely to be misclassifications, thereby improving the overall performance of the model.

\subsection{Probability Flag}\label{subsec:qualityflags}

The final step we employ to improve \textsc{Mummi}'s timescale prediction performance is to define confidence intervals for each probability distribution generated by the ensemble of models. We do this by modeling the width of the peak in the probability distribution and verifying that the probabilities adjacent to the peak are lower than the peak probability by a specified margin. Combined, these two approaches flag galaxies with flat probability distributions for the timescale predictions—those without a clear peak—thus allowing us to exclude such cases from the final pool of predictions.

For each galaxy we first average the probability vectors of the 20 networks and build the associated cumulative distribution function (CDF). We then evaluate a monotonic cubic interpolation of the CDF to estimate the probability contained between the points where the CDF reaches 0.16 and 0.84 around the mode. This $\pm 1\sigma$ interval defines the width of the dominant peak without imposing a Gaussian shape and provides both lower and upper bounds on the peak prediction. We then define a limit for each timescale bin that maximizes performance while balancing completeness.

\begin{figure}
    \centering
    \includegraphics[width=0.95\linewidth]{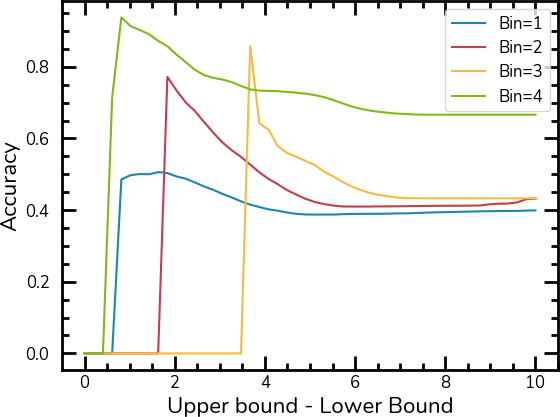}
    \caption{\textbf{Accuracy as a function of snapshot confidence intervals for the \textsc{Mummi} framework}. Each curve represents a specific temporal bin: Bin 1 $(0 
    \ \rm Gyr < T_{PM} < 0.16 \ \rm Gyr)$, Bin 2 $(0.16 \ \rm Gyr < T_{PM} < 0.48 \ \rm Gyr)$, Bin 3 $(0.48 \ \rm Gyr < T_{PM} < 0.96 \ \rm Gyr)$, and Bin 4 $(0.96 \ \rm Gyr < T_{PM} < 1.76 \ \rm Gyr)$. Wider bins capture more generalized temporal trends, resulting in increased classification accuracy for all bins. However, the loss of finer temporal resolution is evident, particularly for Bin 2 and Bin 3, which show declining accuracy for narrow bin widths due to overlapping morphological features across adjacent time intervals. This figure highlights the trade-off between temporal precision and classification accuracy in the MUMMI framework, emphasizing the importance of optimizing bin widths for specific scientific objectives.}
    \label{fig:qualityflags}
\end{figure}

In Figure \ref{fig:qualityflags}, we show the model’s accuracy when predictions are limited to specific confidence intervals (in snapshots) for each timescale bin. As shown, selecting predictions with smaller peak widths yields higher accuracy. The curves drop to 0 at small widths because there are no predictions beyond a certain interval, given that each timescale bin has its own snapshot width.

Additionally, we check whether the peak probability is at least twice that of the adjacent bins. This helps capture distributions that are broad and have long tails yet feature sharp peaks, which might have wide confidence intervals under the previous criterion. Any galaxy meeting one of these two criteria is flagged as having a reliable probability distribution. By selecting only the flagged galaxies, we improve overall performance while mitigating the reduction in completeness caused by the smaller sample. Around $72\%$ of the sample of post-mergers receive this quality flag, hence the majority of the sample is preserved.

\begin{figure}
    \centering
    \includegraphics[width=0.95\linewidth]{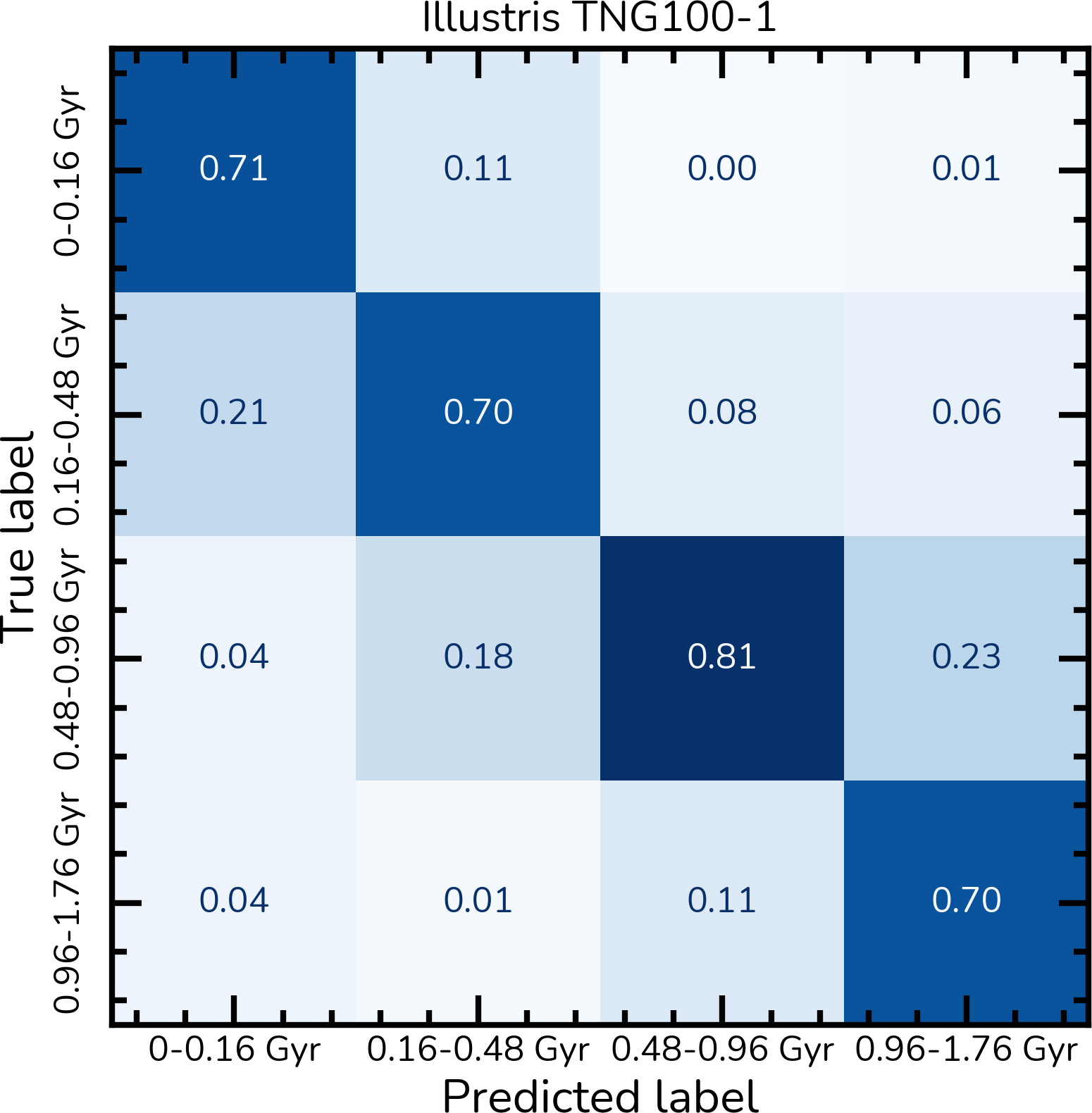}
    \caption{\textbf{Confusion matrix illustrating the performance of the \textsc{Mummi} framework in classifying post-merger galaxies into four temporal bins: $0 < TPM < 0.16 \ Gyr$, $0.16 < TPM < 0.48 \ Gyr$, $0.48 < TPM < 0.96 \ Gyr$, and $0.96 < TPM < 1.76 \ Gyr$}. The diagonal elements represent the fraction of correctly classified galaxies within each bin, with accuracies ranging from 70\% to 81\%. Off-diagonal elements indicate misclassifications, reflecting transitions between adjacent bins due to the continuous nature of galaxy evolution. The highest classification accuracy is observed for the 0.48–0.96 Gyr bin, while early and late post-mergers exhibit modest overlaps with neighbouring bins.}
    \label{fig:cm-small}
\end{figure}

In Figure \ref{fig:cm-small}, we show the confusion matrix for galaxies flagged as having reliable probability distributions. The overall performance rose to between 70\% and 81\%, with improvements of 9\%, 24\%, 33\%, and 2\% in each bin, respectively. The tendency for misclassifications to cluster in adjacent bins remains. This increase in purity enables us to apply the model to observations, provided we assume the same quality-flag criteria.

\subsection{Propagation of contaminants}\label{subsec:contaminants}

{ A critical aspect of finding and characterizing galaxy mergers in the local Universe arises from intrinsic low merger fractions at this regime, with studies reporting fractions below $5\%$ \citep{Bundy2009, Bridge2010, Casteels2014, Robotham2014, fuentes2025}. This hinders the reliability of low purity classifiers, given that non-merging galaxies far outnumber merging galaxies, thus limiting their usability. In \cite{Ferreira2024a} we discuss in detail how the first two steps in \textsc{Mummi} address this issue, reducing $75\%$ of false positive rates when classifying galaxy mergers, providing a very robust framework to select pure and complete samples of mergers.

Due to the hierarchical nature of \textsc{Mummi}, misclassifications in earlier steps can propagate through it.  For example, in the validations of Step~2 in \cite{Ferreira2024a}, we showed that $4\%$ of the galaxies labeled as post-mergers are, in truth, late-stage pre-mergers close to coalescence. Although the lingering false positive rate in Step~3 is minimal, it is important to understand where these fall into Step~3 timescale classifications. Recognizing how \textsc{Mummi} processes misclassifications (both non-mergers and pairs) into the time domain helps us to understand the resulting samples it produces. 

To investigate contamination propagation in Step~3, we draw two batch sizes (2048 galaxies) of samples from \textsc{Mummi}'s Step~1 test set at random with no replacement, only non-mergers and pairs, respectively. Then, we process both datasets through all steps, effectively forcing Step~3 to deal with a full sample of misclassifications. We then track what timescales these galaxies are allocated to. We first discuss the case of non-mergers and then the pairs.

When \textsc{Mummi} is applied to the non-merger sample, 78 galaxies passes Step~1 and Step~2 classification, arriving at Step~3 as false positive post-mergers, and they are preferentially assigned longer timescales, with  4, 11, 29, 34 in each bin, respectively. From these 78 non-mergers, only 6 are assigned a good probability flag and are only classified as either $0 \ \rm Gyr < T_{PM} < 0.16 \ \rm Gyr$ or $0.96 \ \rm Gyr < T_{PM} < 1.76 \ \rm Gyr$, with 1 and 5, respectively. Additionally, while inspecting the probability distributions from the late timescale-bin, the peak of the probability is closer to $1.76 \ \rm Gyr$. Given that \textsc{Mummi} has no other category to put these galaxies, it attempts to assign them the longest timescale possible preferentially, which is consistent with these non-merger galaxies merger history, and their original label from \cite{Ferreira2024a}. If we relax our mass ratio threshold to consider mini mergers ($0.01 < \mu < 0.1$),  47 out of the 78 non-mergers can be re-labeled as post-mergers within our timescale selection window, and thus are not entirely incorrect classifications, but at the border of the domain considered by \textsc{Mummi}.}

\begin{figure}
    \centering
    \includegraphics[width=0.95\linewidth]{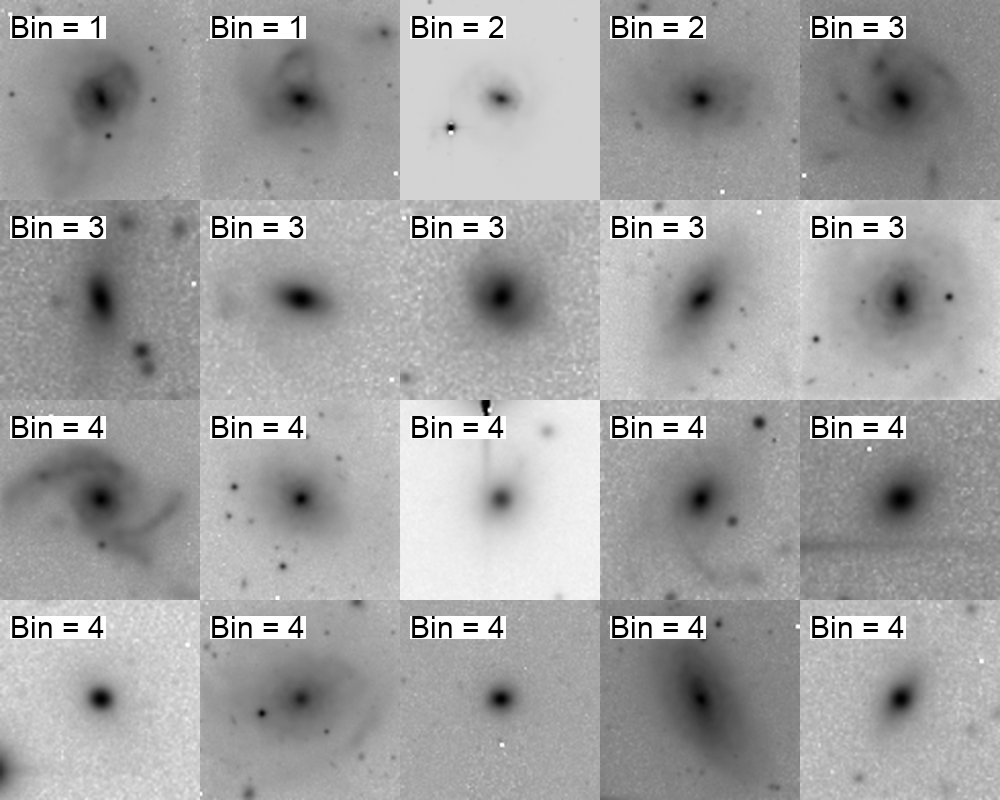}
    \caption{\textbf{Non-mergers classified as Post-mergers}. A random assortment of 20 \textsc{Mummi} non-mergers to post-mergers misclassifications ordered by time the bins. This shows that misclassifications are usually driven by ambiguous morphology, even to the trained eye.}
    \label{fig:mosaic_NM3}
\end{figure}

\begin{figure}
    \centering
    \includegraphics[width=0.95\linewidth]{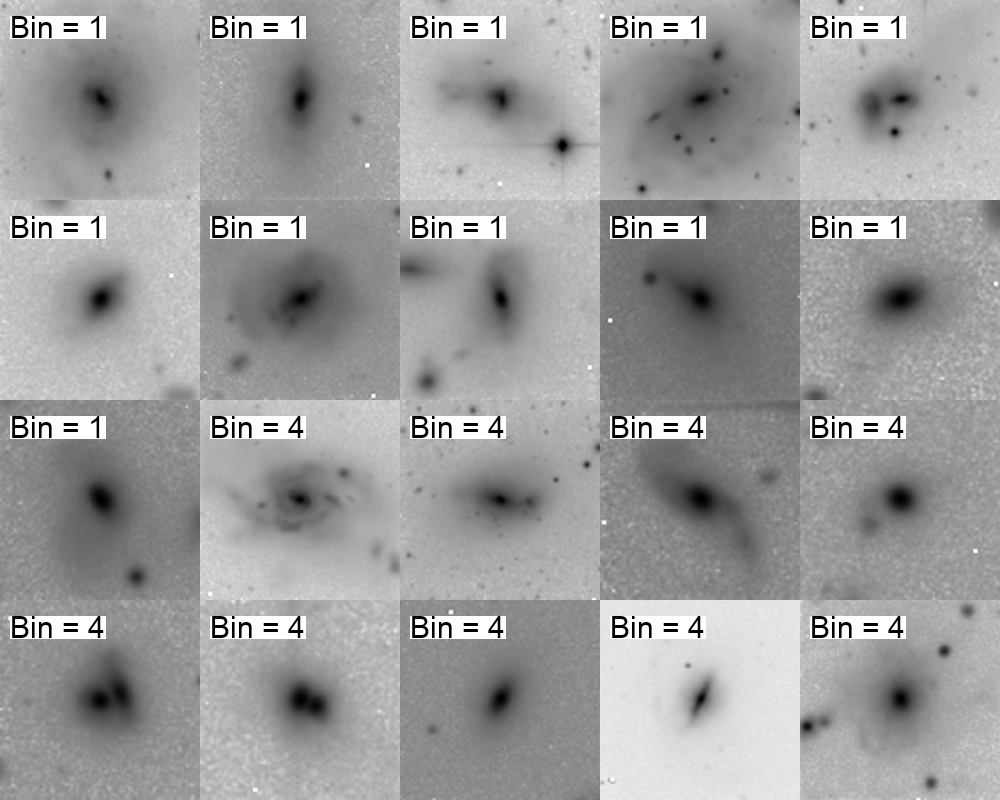}
    \caption{\textbf{Pairs classified as Post-mergers}. A random assortment of 20 \textsc{Mummi} pairs to post-mergers misclassifications ordered by the time bins. This shows that misclassifications are usually driven by ambiguous morphology, even to the trained eye.}
    \label{fig:mosaic_PAIRS}
\end{figure}

{ We also conduct a similar experiment with a sample of 2048 pairs (pre-mergers). This case is slightly different given that galaxies are only classified as pairs in Step~2, thus all the misclassifications entering Step~3 for this test are due to \textsc{Mummi}'s ability to distinguish pairs from post-mergers. In \cite{Ferreira2024a}, we report that \textsc{Mummi} does this with a $\sim97\%$ success rate. Additionally, the imbalance of the problem is mitigated given that our domain is now mergers only. However, here we find that out of the 2048 galaxies, 153 enter Step~3 as pairs classified as post-mergers. The timescale distribution is also very similar to the non-merger case, with misclassifications assigned progressively to longer timescales, with 19, 34, 50, 50, respectively, with only 11 receiving a good probability flag. The 11 flagged galaxies are only classified into $0 \ \rm Gyr < T_{PM} < 0.16 \ \rm Gyr$ or $0.96 \ \rm Gyr < T_{PM} < 1.76 \ \rm Gyr$ bins, with 4 and 7 galaxies, respectively.  Investigating their merging histories based on lower mass ratios ($\mu < 0.1$) we find that 41 out of 153 could be re-labeled as post-mergers within our time window, thus not being entirely incorrect but slightly beyond our classification criteria. 

In Figure~\ref{fig:mosaic_NM3} and Figure~\ref{fig:mosaic_PAIRS} we show a random selection of 20 non-merger and pair galaxies from this experiment, respectively. It is clear that galaxies misclassified by \textsc{Mummi} are not obvious mistakes, showcasing the challenge of defining the appearance of some of these objects. Most of the misclassified galaxies display clear disruption-like features in the immediate bins, and fuzzy outskirts in the longer timescale ones, and many with projected companions that could be disguised as elongated features and/or are embedded in dense environments. The galaxies in these images are confusing objects even to the trained eye, and could represent a more subtle merger history involving lower mass ratio interactions (mini-mergers), or mergers interacting on even longer timescales ($> 1.76 \ \rm Gyr$) that are not tracked by our labeling \citep{Bottrell2024, byrnemamahit2025}.

Overall, misclassifications propagated from earlier steps in the \textsc{Mummi} pipeline are preferentially assigned to long timescales before the probability flags. The ones that remain after the flag filter are then the cases that can be misinterpreted as very disturbed due to recent coalescence, or that present subtle faded-out features or companions on longer timescales. Moreover, the addition of the flag essentially reduces the rate of false positives in Step~3 ten-fold, getting rid of ambiguous cases where time inference is unclear. The resulting samples of positive cases that stem from the full pipeline are always usually small (a few hundred) and on the realm of possibility for visual inspection for final veto.}

\subsection{Star Formation Rates Biases with Sample Selection}\label{subsec:bias}

The strict selection criteria employed to boost model performance (\S \ref{subsec:qualityflags}) might raise concerns regarding biases that could impact the final sample of post-mergers when we move to real UNIONS galaxies. For example, are we preferentially identifying galaxies with particular characteristics? To investigate whether our methodology has any impact on assessment of physical characteristics of post-mergers thus selected, we explore whether the increase in purity (and decrease in completeness) affects the star formation rates of the resulting sample of post-merger regime in TNG galaxies. We choose to conduct this experiment with the SFR as it is one of the physical properties not part of the initial selection of our sample.

\begin{figure*}
    \centering
    \includegraphics[width=0.95\linewidth]{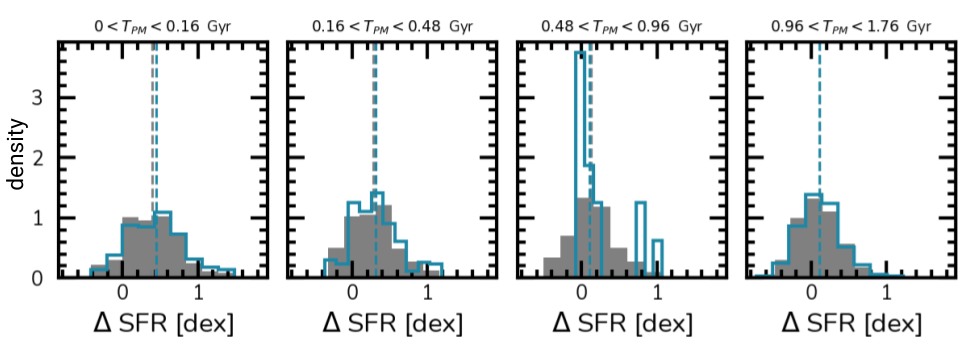}
    \caption{\textbf{Distribution of $\Delta \rm SFR$ for post-merger galaxies in four time bins after coalescence ($0 < T_{PM} < 0.16 \ Gyr$, $0.16 < T_{PM} < 0.48 \ Gyr$, $0.48 < T_{PM} < 0.96 \ Gyr$, and $0.96 < T_{PM} < 1.76 \ Gyr$)}. Each panel shows the histogram of $\Delta \rm SFR$ for the whole TNG sample in grey while blue histogram outlines the selection with the quality flags applied upon to increase purity. The corresponding medians of each distribution are outlined with vertical dashed lines illustrate illustrating the agreement between the parent sample and the purity-enhanced sample. Note that the dashed gray line is perfectly behind the blue dashed line in the last bin histogram plot.}
    \label{fig:ssfr}
\end{figure*}

First, we define the star-forming main sequence (SFMS) from the fits in \cite{Speagle2014} for all galaxies in our sample. We follow the \cite{Speagle2014} treatment because it matches the SFMS in IllustrisTNG over the stellar-mass and redshift ranges explored here \citep{Donnari2019}. Then, we estimate the difference between the instantaneous SFR for each galaxy and how it deviates from the SFMS from \citep{Speagle2014} ($\Delta SFR$) in each timescale bin. Finally, we compare the distributions of $\Delta SFR$ for galaxies with and without the quality flag quality criteria. In Figure \ref{fig:ssfr}, we illustrate the distribution of $\rm \Delta SFR$ shown as a blue histogram when the probability flag is applied, and compare it to the true distribution of labels from the simulation (grey histogram) without any cuts. The vertical dashed lines in matching colours indicate the median values for each distribution.

All distributions and medians are in close agreement, except for the $0.48 < T_{PM} < 0.96\,\mathrm{Gyr}$ bin in the third panel, which exhibits poor sampling due to a small sample size (low completeness). Thus, using the probability flag should not introduce any significant selection bias in the samples produced by \textsc{Mummi}, and the final resulting samples are representative from the parent full sample. 

For the actual time evolution of $\rm \Delta SFR$ in UNIONS galaxies, see \cite{Ferreira2024b}, where we explore the evolution of $SFR$ enhancements and how that plays a role into merger-triggered mass growth in galaxies throughout the post-merger sequence.

\subsection{Post-merger timescales in UNIONS}
\label{subsec:UNIONSresults}

The initial goal of \textsc{Mummi} was to find, classify and extract important information about mergers from the UNIONS survey. Following \citet{Ferreira2024a}, where \textsc{Mummi} generated a large catalog of mergers (simple majority or unanimous) from the UNIONS DR5 and SDSS DR7 overlapping region—yielding 12,103 (3,847) post-mergers that received at least 11 (20) votes in Step 1—we now apply the new Step 3 to measure post-merger timescales. From these, 8,716 (2,272) pass the probability quality-flag criteria and yield well-behaved probability distributions. The breakdown of the sample across different timescales is as follows: 924 (475), 808 (405), 825 (407), and 6,159 (905) for $0 < T_{PM} < 0.16\,\mathrm{Gyr}$, $0.16 < T_{PM} < 0.48\,\mathrm{Gyr}$, $0.48 < T_{PM} < 0.96\,\mathrm{Gyr}$, and $0.96 < T_{PM} < 1.76\,\mathrm{Gyr}$, respectively.

\begin{figure*}
    \centering
    \includegraphics[width=0.95\linewidth]{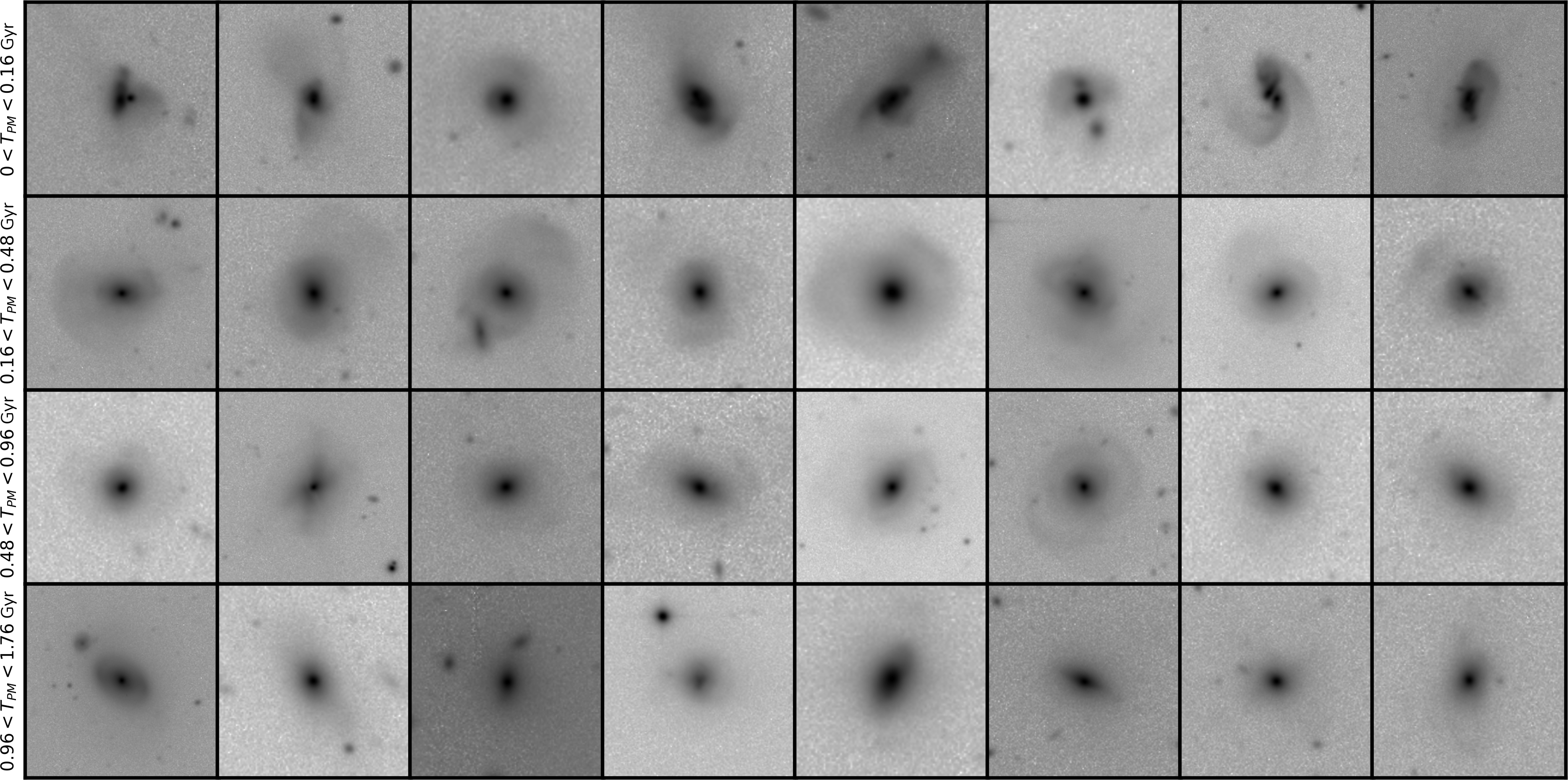}
    \caption{\textbf{A mosaic of post-merger galaxies from the UNIONS survey classified into four temporal bins by the \textsc{Mummi} framework}. The rows represent increasing time since coalescence, with early post-mergers exhibiting prominent morphological disturbances such as tidal tails and asymmetries, gradually transitioning to more relaxed morphologies in later bins.}
    \label{fig:UNIONS_mosaic}
\end{figure*}

In Figure \ref{fig:UNIONS_mosaic}, we show 32 UNIONS galaxies, with each row corresponding to a different timescale bin, increasing from top to bottom. These galaxies were randomly selected from the classified sample. Figure \ref{fig:UNIONS_mosaic} illustrates the clear transitions over time following a merger event: galaxies in the immediate timescale show strong tidal tails and incompletely coalesced central stellar masses, whereas those at short timescales exhibit prominent shells and streams. Although similar features persist in the intermediate bin, they appear more faded. Finally, galaxies at long timescales are characterized by a lack of strong features, displaying only minor signatures of disturbances that have likely faded over the last Gyr.

These robust statistics, even with unanimous selection (for increased purity), yield large populations with which to track physical processes across the post-merger regime, allowing for further targeted selections. For example, in \citet{Ferreira2024b}, this UNIONS post-merger timescale sample was used with BPT cuts to select only star-forming galaxies, illustrating the evolution of star-formation enhancements in star-forming post-mergers. Additionally, \citet{Ellison2024b} used the same catalog, cross-matched with post-starburst selections, to investigate the time evolution of quenching in post-mergers. Moreover, \citet{Ellison2025} matched these time predictions with AGN classifications, demonstrating the excess of AGN in post-mergers and their time dependence. Finally, \citet{Ellison2026} using the time slicing from \textsc{Mummi} showed that post-merger galaxies have $10$ to $20\%$ excess stellar mass in their central regions when compared to non-mergers of similar mass and redshift.

\section{SUMMARY}
\label{sec:summary}
In this work, we extended the capabilities of the \textsc{Mummi} framework to predict post-merger timescales for galaxies, utilizing a combination of convolutional neural networks and vision transformers. By leveraging mock observations derived from the IllustrisTNG simulations, we trained an ensemble of models to classify galaxies into four temporal bins spanning 0 to 1.76 Gyr after coalescence. This approach achieved robust classification accuracies, exceeding 70\% in all bins, with the highest accuracy observed in the intermediate time bin (0.48–0.96 Gyr). The results demonstrate the ability of \textsc{Mummi} to distinguish distinct evolutionary phases within the post-merger regime.

The application of this methodology to the UNIONS survey has yielded a catalog of 8,716  post-merger galaxies, providing temporal classifications alongside physical properties such as stellar mass and redshift, which is publicly available\footnote{Catalogs can be found at \url{https://github.com/astroferreira/MUMMI_UNIONS}}.

These data also enabled new insights into the temporal evolution of key processes, including star formation enhancements \citep{Ferreira2024b}, quenching \citep{Ellison2024b}, AGN activity \citep{Ellison2025}, and stellar mass build-up \citep{Ellison2026} over the post-merger timeline. This study highlights the potential of combining advanced machine learning techniques with simulation-based datasets to address long-standing challenges in galaxy evolution. Future work will focus on extending the framework to higher redshifts, in ground-based and space-based observations, as well as into the pre-merger phase, where similar techniques can be potentially applied to extract temporal information and mass ratios of the pre-coalescence stage.

\section*{Acknowledgements}
\addcontentsline{toc}{section}{Acknowledgements}
We are very thankful to the anonymous referee whose comments improved this paper greatly. We are honored and grateful for the opportunity of observing the Universe from Maunakea and Haleakala, which both have cultural, historical and natural significance in Hawaii. This work is based on data obtained as part of the Canada-France Imaging Survey, a CFHT large program of the National Research Council of Canada and the French Centre National de la Recherche Scientifique. Based on observations obtained with MegaPrime/MegaCam, a joint project of CFHT and CEA Saclay, at the Canada-France-Hawaii Telescope (CFHT) which is operated by the National Research Council (NRC) of Canada, the Institut National des Science de l’Univers (INSU) of the Centre National de la Recherche Scientifique (CNRS) of France, and the University of Hawaii. This research used the facilities of the Canadian Astronomy Data Centre operated by the National Research Council of Canada with the support of the Canadian Space Agency. This research is based in part on data collected at Subaru Telescope, which is operated by the National Astronomical Observatory of Japan.
Pan-STARRS is a project of the Institute for Astronomy of the University of Hawaii, and is supported by the NASA SSO Near Earth Observation Program under grants 80NSSC18K0971, NNX14AM74G, NNX12AR65G, NNX13AQ47G, NNX08AR22G, 80NSSC21K1572 and by the State of Hawaii.
We acknowledge the support of the Digital Research Alliance of Canada for providing the compute infrastructure to train the models presented here. We thank Sarah Huber for helpful technical support on the usage of the Cedar cluster, troubleshooting and helping with optimizations necessary to manage the data scale of our models and datasets swiftly. We acknowledge and thank the IllustrisTNG collaboration for
providing public access to data from the TNG simulations. SLE, DRP and LF gratefully acknowledge NSERC of Canada for Discovery Grants which helped to fund this research. SJB and SW acknowledge graduate fellowships funding from the Natural Sciences and Engineering Research Council of Canada (NSERC); Cette recherche a été financée par le Conseil de recherches en sciences naturelles et en génie du Canada (CRSNG). LF also gratefully acknowledge funding from CNPq's Conhecimento Brasil program (447108/2024-7).

%%%%%%%%%%%%%%%%%%%%%%%%%%%%%%%%%%%%%%%%%%%%%%%%%%

\section*{Data Availability}

The merger classification catalogs used in this article are publicly available at \url{https://github.com/astroferreira/MUMMI_UNIONS}. A subset of the raw data underlying this article are publicly available via the Canadian Astronomical DataCenter at \url{http://www.cadc-ccda.hia-iha.nrc-cnrc.gc.ca/en/megapipe/}. The remaining raw data and all processed data are available to members of the Canadian and French communities via reasonable requests to the principal investigators of the Canada–France Imaging Survey, Alan McConnachie and Jean-Charles Cuillandre. All data will be publicly available to the international community at the end of the proprietary period, scheduled for 2026.



%\bibliographystyle{mnras}

\bibliographystyle{mnras}
\bibliography{example}

%%%%%%%%%%%%%%%%%%%%%%%%%%%%%%%%%%%%%%%%%%%%%%%%%%

%%%%%%%%%%%%%%%%% APPENDICES %%%%%%%%%%%%%%%%%%%%%

%%%%%%%%%%%%%%%%%%%%%%%%%%%%%%%%%%%%%%%%%%%%%%%%%%

% Don't change these lines
\bsp	% typesetting comment
\label{lastpage}
\end{document}